\documentclass[preprint]{aastex631}

\usepackage{graphicx}
\begin{document}
\title{M33 Cepheids from CFHT/MegaCAM survey}

\author{Samuel Adair}
\affiliation{University of Hawaii at Hilo \\
Hilo, HI 96740, USA}

\author{Chien-Hsiu Lee}
\affiliation{W. M. Keck Observatory \\
Kamuela, HI 96743, USA}

\begin{abstract}

In this paper we analyze Sloan g,r,i archival imaging data 
of M33 taken by Hartman et al. (2006) using Megacam at the Canada-France-Hawaii Telescope. To determine the distance to the M33 galaxy, we performed several analytical steps to identify its Cepheid population. We used the Lomb-Scargle algorithm to find periodicity and visually identified 1989 periodic variable stars. 
Since Cepheids occupy a specific region of the color-magnitude diagram, to differentiate Cepheids from other variables we used the expected position of the Cepheid instability strip to down-select Cepheids in M33 from other variables. This led to our sample of 1622 variables, the largest Cepheid sample known in M33 to date. We further classified these Cepheids into different sub-classes, and used the fundamental mode Cepheids to estimate distance moduli for M33 in different filters:
\textmu = 25.044 \textpm 0.083 mag in the g filter, \textmu = 24.886 \textpm 0.074 mag in the r filter, and \textmu = 24.785 \textpm 0.068 mag for the i filter.
These results are in agreement with previous results. 
\end{abstract}



\section{Introduction}
In the era of precision cosmology, an instrumental approach to test cosmological models is to compare the Hubble constant determined from different methods, especially from locally and at the early Universe. There has been efforts on both fronts, for example, from the distance ladder methods (\cite{2021arXiv211204510R} using Cepheids, and \cite{2021ApJ...919...16F} using TRGB stars), the strong lensing time-delays \citep{2020MNRAS.498.1420W}, and results from CMB experiments such as Planck \citep{2020A&A...641A...6P}. There are various discrepancies among the results, pointing to possible underestimated systematics in the methods themselves or hinting of new physics.   

As one of the closest spiral galaxies in the Local Group, M33 provides tremendous leverage to study its stellar content. Its proximity allows us to resolve brighter variable star populations with ground-based telescopes. With a simple geometry and as a face-on galaxy, we can treat all the M33 stars as at the same distance. In addition, M33 is a metal rich environment compared to the Magellanic Clouds and other dwarf galaxies, and can be used to study metallicity influence on the properties of variables \citep[see e.g.][]{2009MNRAS.396.1287S}.

Previous experiments have revealed tens to hundreds of Cepheids in M33 and used them to derive the distance to M33. \cite{1983ApJ...267L..25S} derived a apparent blue distance modulus of 25.35 mag by analyzing blue light curves and periods between 37 and 3 days for 13 Cepheids. \cite{1991ApJ...372..455F} derived a distance modulus of 24.64 $\pm$ 0.09 mag based on BVRI CCD photometry. The DIRECT project used the 1m FLWO telescope and presented 251 Cepheids in M33 \citep{2001AJ....121..870M}. The Araucaria project \citep{2013ApJ...773...69G} have presented detailed infrared observations of 26 Cepheids discovered by the DIRECT project with VLT and inferred a distance modulus of 24.62 $\pm$ 0.07 mag. \cite{2011ApJS..193...26P} have augmented the DIRECT project with longer time-span with WIYN/ODI observations and presented a sample of 564 Cepheids. Using this sample, they derived a distance modulus of 24.76 $\pm$ 0.02 mag after quantifying biases in the photometry due to the crowding effect. In addition, \cite{2006ApJ...653L.101B} used archival M33 data from CFHT/MegaCAM and presented a sample of beat Cepheids to study the metallicity gradient of M33. However, the Cepheid distance to M33 remains controversial and provided a smaller distance estimate when compared to other methods. 
In this work, we perform a detailed study of the Cepheid population in M33 with archival CFHT/MegaCAM data \cite{2006MNRAS.371.1405H}, and present the largest Cepheid sample up-to-date. This paper is organized as follows: in section \ref{sec.data} we briefly summarize the CFHT/MegaCAM M33 data-set. In section \ref{sec.analysis} we provide details of our analysis method to identify and classify Cepheids. We compare our sample with previous results in section \ref{sec.results}, followed by a conclusion in section \ref{sec.conclusion}. 

\section{Data}
\label{sec.data}
To investigate the Cepheids in M33, we utilized the archival time-series observations with CFHT by \cite{2006MNRAS.371.1405H}. An 1-$deg^2$ region centered at M33 was monitored repeatedly with the  wide-imager MegaCAM. Data was taken on 27 nights between 2003 - 2005 in three optical filters $g$, $r$, and $i$. Each exposure was $\sim$ 500 seconds in $g$ and $\sim$ 600 seconds in $r$ and $i$, with a median seeing of 0.95",0.85",0.77" in $g$, $r$, and $i$, respectively. 

The raw data were reduced with the Elexir pipeline to correct for bias, dark, flat-field, and fringing. Since M33 is a very crowded field, \cite{2006MNRAS.371.1405H} deployed the image subtraction method devised by Alard \& Lupton (2000) to find variable sources down to the limiting magnitude of individual images. This led to a catalog of more than 36000 point sources with variability within the time-span of the CFHT M33 monitoring program. However, \cite{2006MNRAS.371.1405H} did not present further classification or periodicity of these variable sources, so we conduct our own analysis shown in the following section.

\section{Analysis}
\label{sec.analysis}
We started with more than 36000 variable sources from \cite{2006MNRAS.371.1405H}. To reduce the number of variable sources, we only analyzed sources that showed variability and had light curves in all three filters, reducing the number of variable sources to 23309. To ensure sufficient number of data points in the light curves for a period search, we ruled out the variable sources that had less than 10 data points in each filter, reducing the number of variable sources to 22539. For the variable sources that we did not rule out, we used the fast Lomb-Scargle periodogram algorithm implemented by \cite{2015ApJ...812...18V} to determine the best period in each filter. 

With the periods in hand, we then produced phase-folded light curves and visually examined them.
We used the shape of the light curve to visually determine whether the variable sources showed reliable periodicity such that we could use them for further analysis. We note there might be a bias to remove noisy light curves with visual inspection, but this approach ensured that all the light curves had reliable periods as vetted by eye.
Visual determination led to 1989 variables with periodicity. To determine the best period for each variable source, we calculated the ratio between the periods in the g and r filters, the g and i filters, and the r and i filters. If the ratio was within 20 percent and the period was greater than 1.1, we determined that period was the best period for the corresponding variable source. With the best period, we can start classifying the variable sources as either Cepheids or not.
\subsection{Fourier Decomposition}
\begin{figure}[h!]
    \centering
    \begin{tabular}{c}
        \includegraphics[width = 12cm]{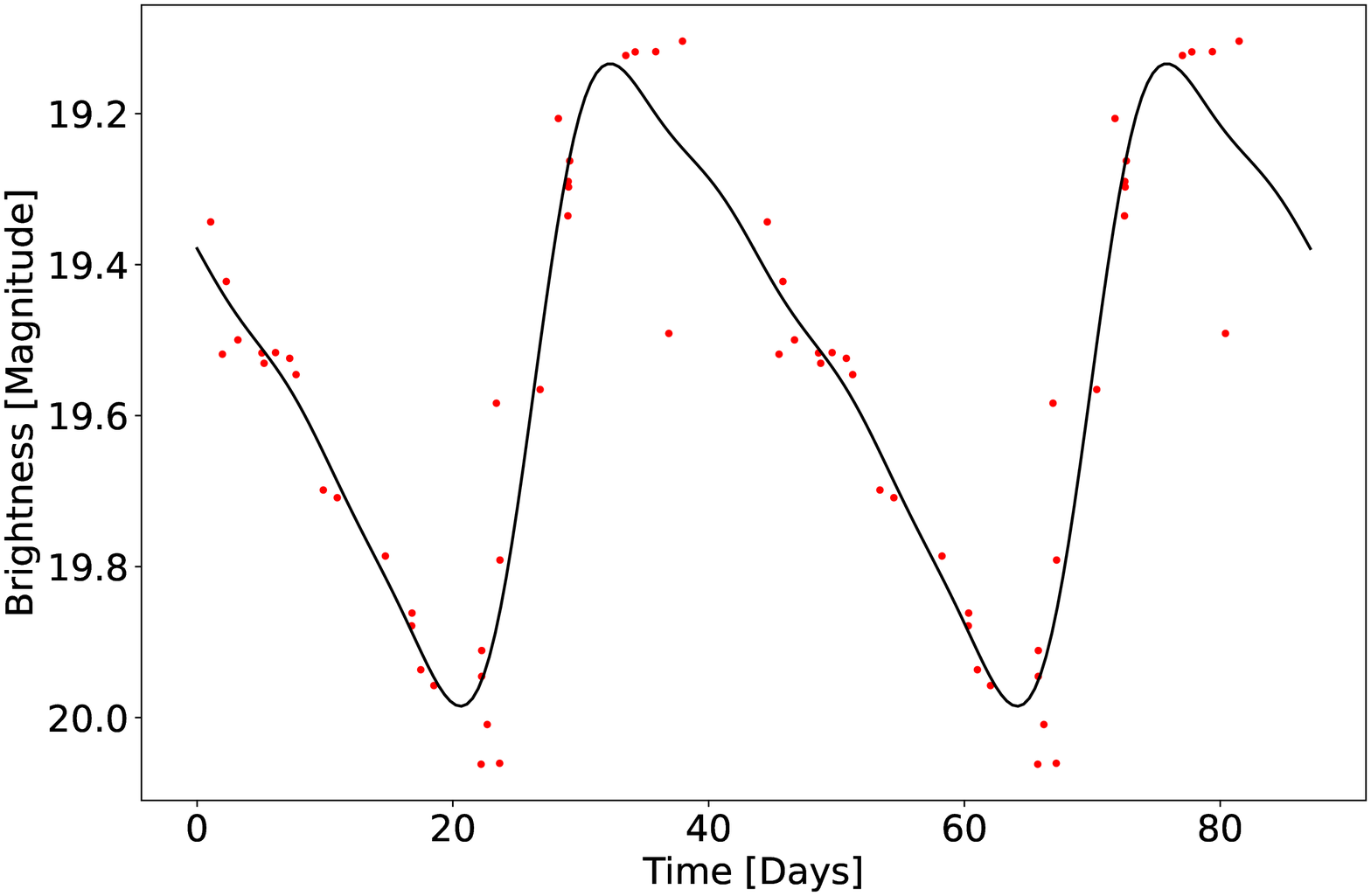} 
    \end{tabular}
\centering
    \begin{tabular}{c}
        \includegraphics[width=12cm]{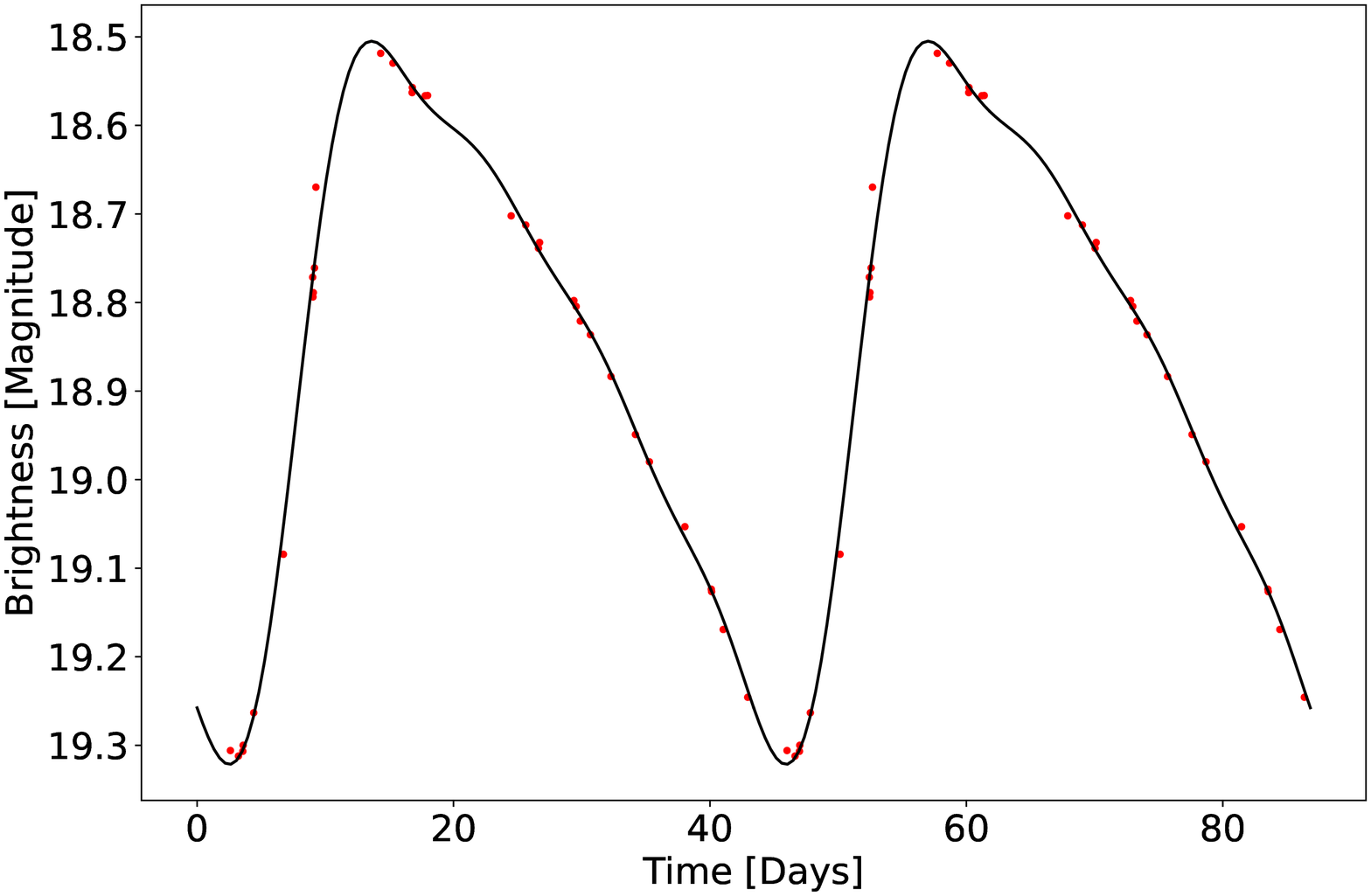}
    \end{tabular}
\centering
    \begin{tabular}{c}
        \includegraphics[width=12cm]{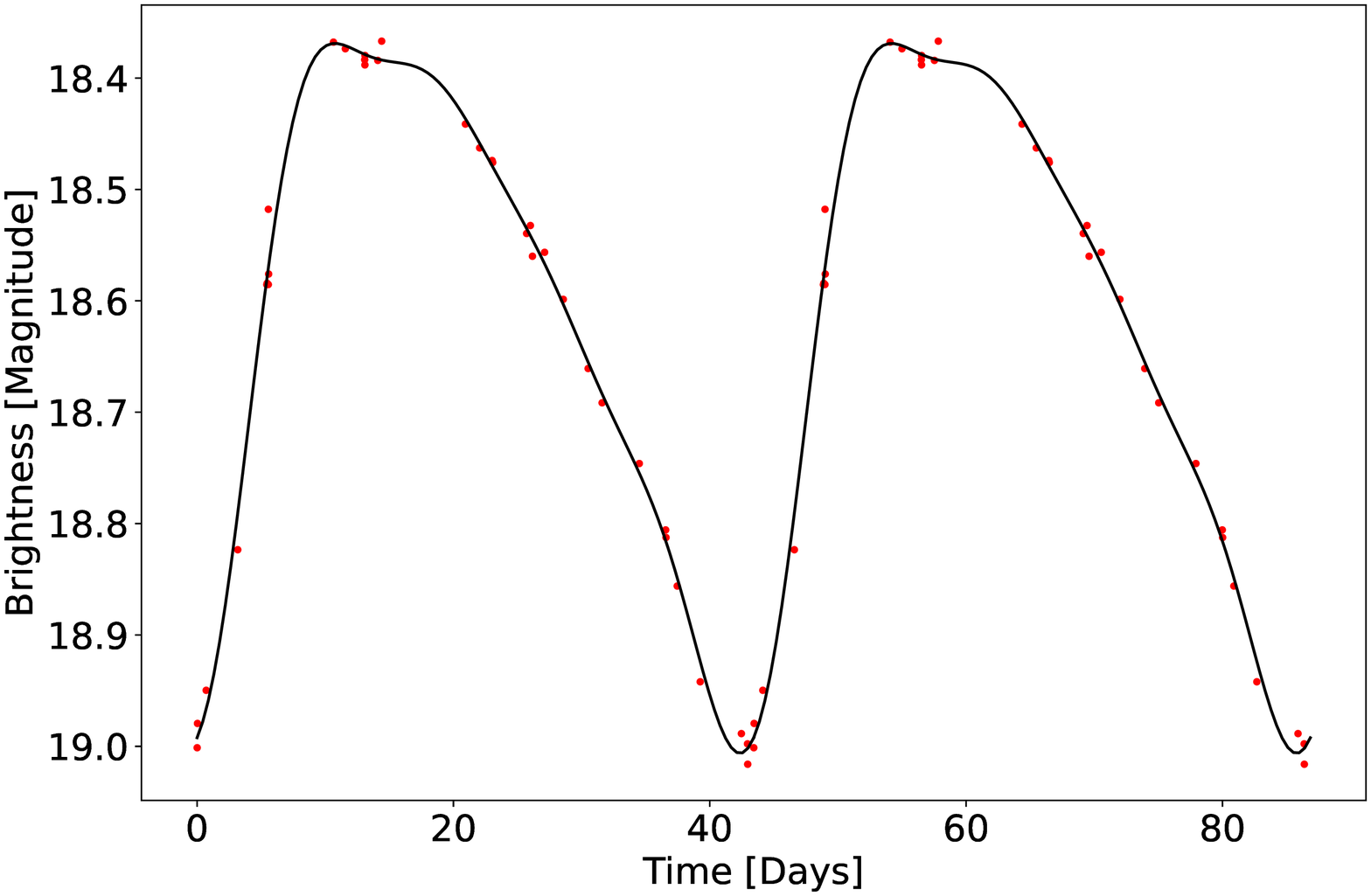}
    \end{tabular}
    \caption{An example of light curves from 30308 with Fourier decomposition in the g, r, and i filters.}
    \label{fig:my_label}
\end{figure}

Since the CFHT M33 light curves are relatively sparse, a practical way to describe the light curve structure is through Fourier decomposition. A Fourier decomposition of the light curve can be used to characterize the variable source and determine whether it is a Cepheid or other type of variable. The Fourier decomposition can also be used to measure the mean magnitude in each filter, and determine the type of Cepheid. For our Fourier decomposition, we used the equation

\begin{equation}
    m(x) = a_{0} + \sum_{n=1}^{N} a_{n}cos((2\pi)nx) + b_{n}sin((2\pi)nx)
\end{equation}

with x being the modified Julian date divided by the period. We use the Fourier decomposition with N=4 for every variable source and in all three filters. Examples of light curves with Fourier decomposition can be found in Fig. \ref{fig:my_label}.

We can use the coefficients of the Fourier decomposition to determine the mean magnitudes in different filters and calculate the extinction free Wesenheit magnitude as follows. 
\subsection{Wesenheit magnitude}
\begin{figure}[h!]
\centering
\includegraphics[width=12cm]{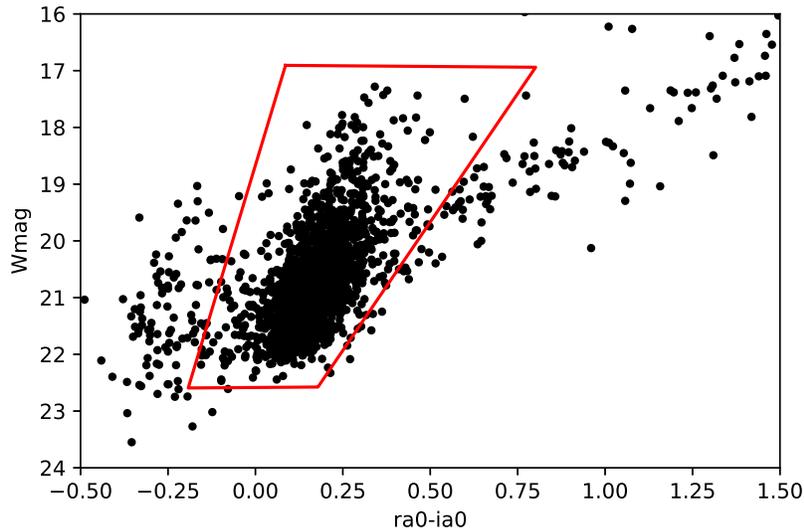}
\caption{This plot shows the color-magnitude diagram of our visually examined periodic variables in the Wesenheit magnitude v.s. r-i color. The red region indicates the instability strip where we identify 1622 Cepheids.}
\label{fig:1}
\end{figure}

With a given dust extinction law, we can use a combination of magnitudes and colours to derive an extinction free magnitude, so-called the Wesenheit magnitude \citep[see e.g.][]{1982ApJ...253..575M}.
We derived the Wesenheit magnitude using the following equation: 
\begin{equation}
W = m_{1} \mathrm{-} R( m_{1} \mathrm{-} m_{2}),
\end{equation}
where $m_{1}$ is the mean magnitude $a_{0}$ from the Fourier decomposition in one filter, $m_{2}$ is the mean magnitude $a_{0}$ from another filter, and the color term factor R can be calculated by the equation:
\begin{equation}
R = \frac{A_{1}}{A_{1}\mathrm{-}A_{2}},
\end{equation}
With $A_{1}$ and $A_{2}$ being the reddening factor corresponding with filters $m_{1}$ and $m_{2}$. In our case, we used the $r$ and $i$ filters to calculate R. We used the $A_{r}$ and $A_{i}$ values from \cite{2011ApJ...737..103S}. Since the CFHT filters have central wavelength close to DES-r and DES-i, we adopted an $A_{r}$ of 2.176 and $A_{i}$ of 1.595 assuming a Galactic extinction law with $R_{V}$=3.1. This led to a color term factor of R = 3.75. With the Wesenheit magnitude in hand, 

we can locate the instability strip region in the color magnitude diagram defined by Wesenheit magnitude from $r$ and $i$ filters vs $r$-$i$ color. Specifically, we adopted the instability strip region given in \cite{2013AJ....145..106K} and adjusted the magnitude boundaries accordingly given the distance difference between M31 and M33 (see Fig. \ref{fig:1}). This led to a sample of 1622 Cepheids.
We referenced our cut to the cut used in \cite{2013AJ....145..106K}. The cut \cite{2013AJ....145..106K} used, and so as we did, is relaxed from the Classical Cepheid instability strip so we do not miss Cepheids with poor photometric measurements. In doing so, we were using cuts bluer than the Classical Cepheid instability strip, hence there are type II Cepheids in our sample.

\subsection{Type Classification}
With the Cepheids identified from the color-magnitude diagram in our data set, we further classified Cepheids into different types. To separate the First Overtone type Cepheids from the Fundamental Mode type Cepheids, we visually placed a line short ward in log(Period) from the main PL relation and chose everything above it as being overtones. 

\begin{equation}
    P < 6.0\mathrm{days} \quad and \quad A_{21} < -0.498 log(P)+0.528
\end{equation}

\begin{figure}[h!]
\centering
\includegraphics[width=12cm]{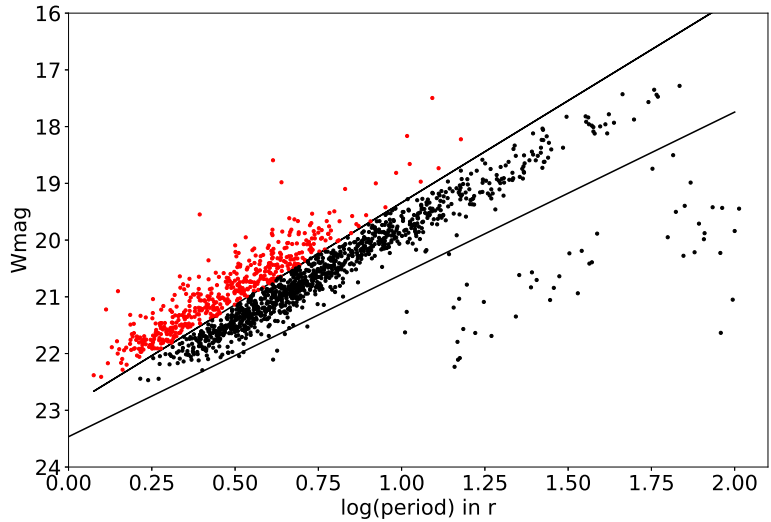}
\caption{Wesenheit magnitude v.s. log(P) of our 1,622 Cepheids identified from the instability strip region. The red points are FO Cepheids classified by the amplitude ratio. The black line is a visual cut to identify the II Cepheids.}
\label{fig:4}
\end{figure}

We also attempted to use the phase difference as in \cite{2013AJ....145..106K} to determine the type II Cepheids but it proved to not be effective for our data set. Since type II Cepheids are known to be less luminous and distribute in a lower luminosity and longer period region on the period-luminosity parameter space, we can visually characterize type II Cepheids using a cut of (see Fig.\ref{fig:4}):
\begin{equation}
    r-band \quad magnitude < -2.29log(P) + 23.8
\end{equation}

\begin{figure}[h!]
\centering
\includegraphics[width=12cm]{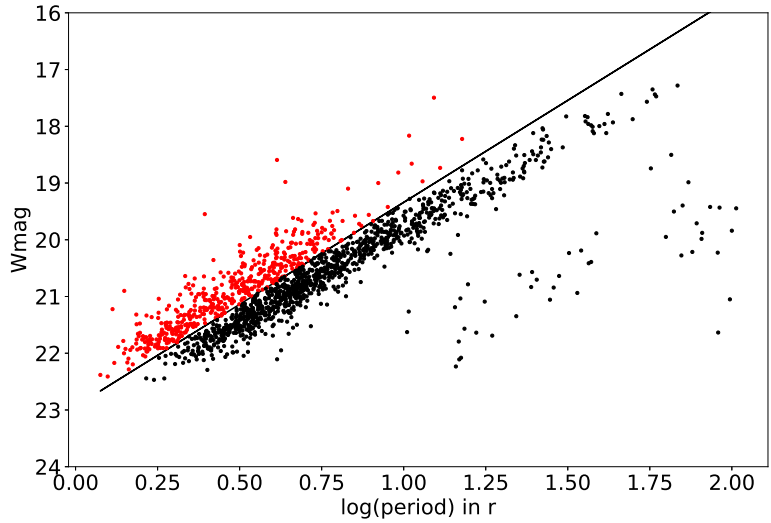}
\caption{Distributions of amplitude ratio v.s. log (Period) from r-filter. The red line shows the boundary to separate the FO type Cepheids from the FM type and type II Cepheids, where the FO type Cepheids are located under the line.}  
\label{fig:3}
\end{figure}

With all the Cepheids characterized by type, we have a clean sample of 1,087 FM Cepheids that are suitable for distance estimation. Table \ref{tab:1} shows the methods in each step of our analysis and the numbers of sources in each step.  

\begin{table}[h!]
    \centering
    \begin{tabular}{|c|c|c|}
        \hline
        Band & Selection criteria & Number of remaining sources \\
        \hline
        g,r,i & data in all three filters & 23,309 \\
        \hline
        g,r,i & less than 10 data points & 22,539 \\
        \hline
        g,r,i & visual light curve & 1,989 \\
        \hline
        r,i & instability strip domain & 1,622 \\
        \hline
        r & FM type Cepheids & 1,087 \\
        \hline
        r & FO type Cepheids & 489  \\
        \hline
        r & Type II Cepheids & 46 \\
\hline
    \end{tabular}
    \caption{Cepheid selection criteria and summary of sources}
    \label{tab:1}
\end{table}

\section{Results}
\label{sec.results}
We have a large sample of 1622 Cepheids, the largest sample collected from M33 to date. There were 1087 FM type Cepheids, 489 FO type Cepheids, and 46 type II Cepheids. In table \ref{tab:2}, we can see a small portion of our Cepheid catalogue. 
\begin{table}[h!]
\centering
    \begin{tabular}{|c|c|c|c|c|c|c|c|c|c|c|c|}
        \hline
Identifier & RA & Dec & g-Per & r-Per & i-Per & g mag & r mag & i mag & A$_{21}$ & W mag & Type \\
        \hline
20018 & 23.790375 & 31.089333 & 2.8107 & 2.8106 & 2.8113 & 22.667 & 22.291 & 22.110 & 0.46038 & 21.612 & FM \\
        \hline
20019 &	23.790458 & 31.008500	 & 2.0543 & 2.0542 & 2.0540 & 22.683 & 22.383 & 22.249 & 0.42264 & 21.884 & FM \\
        \hline
20048 &	23.803958 & 30.945361 & 2.0446 & 2.0448 & 2.0448 & 22.239 & 21.941 & 21.896 & 0.11577 & 21.772 & FO \\
        \hline
20057 & 23.807958 & 31.026444 & 1.8924 & 1.8927 & 1.8927 & 22.089 & 21.812 & 21.743 & 0.22702 & 21.552 & FO \\
        \hline
20135 &	23.856333 & 30.989527 & 1.7954 & 1.7955 & 1.7956 & 22.977 & 22.685 & 22.524 & 0.50223 & 22.082 & FM \\
        \hline
    \end{tabular}
\caption{Parameters of the Cepheids identified by this work. The columns are: 1. light curve identifier from Hartman et al. (2006); 2. right ascension (J2000) in degrees; 3. declination (J2000) in degrees; 4. period from g-band; 5. period from r-band; 6. period from i-band; 7-9 mean magnitudes in g, r, and i-filters; 10. amplitude ratio from r-band; 11. Wesenheit magnitude from r and i-band; 12. type classification. We only show the first five rows as an example. The full table will be released as online material when the paper is accepted.}
    \label{tab:2}
\end{table} 

\begin{figure}[h!]
    \includegraphics[width=0.5\linewidth]{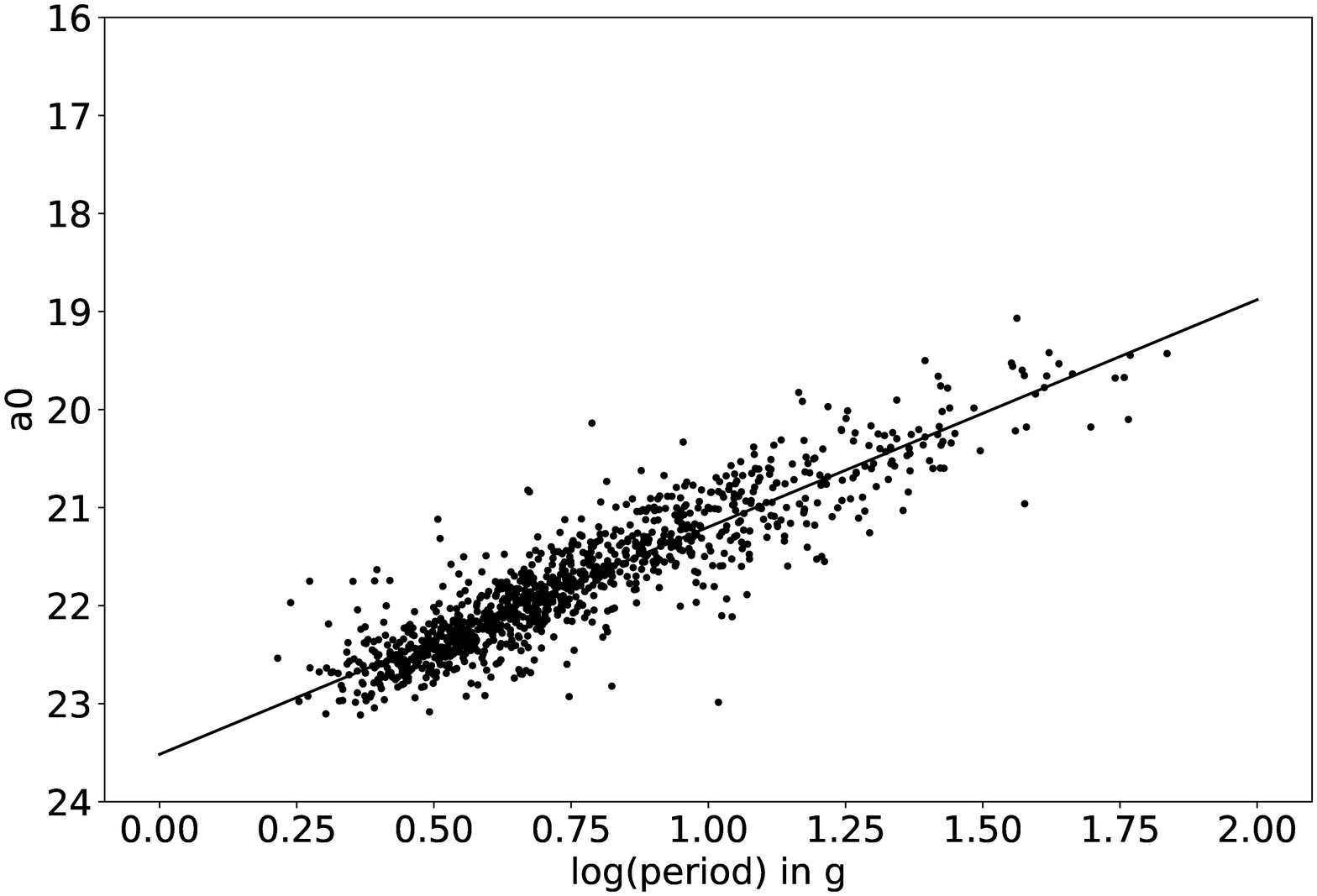}
    \includegraphics[width=0.5\linewidth]{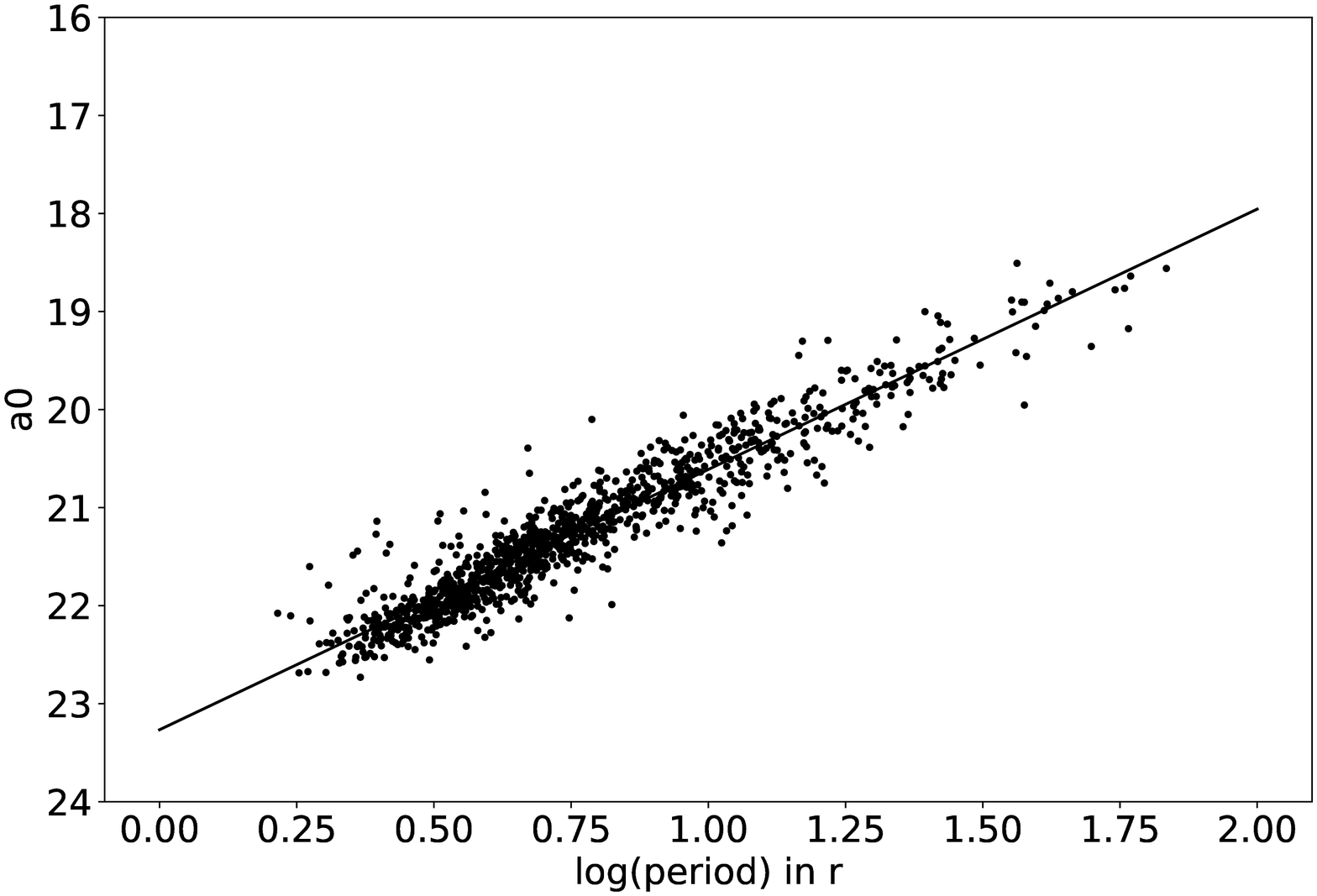}
    \includegraphics[width=0.5\linewidth]{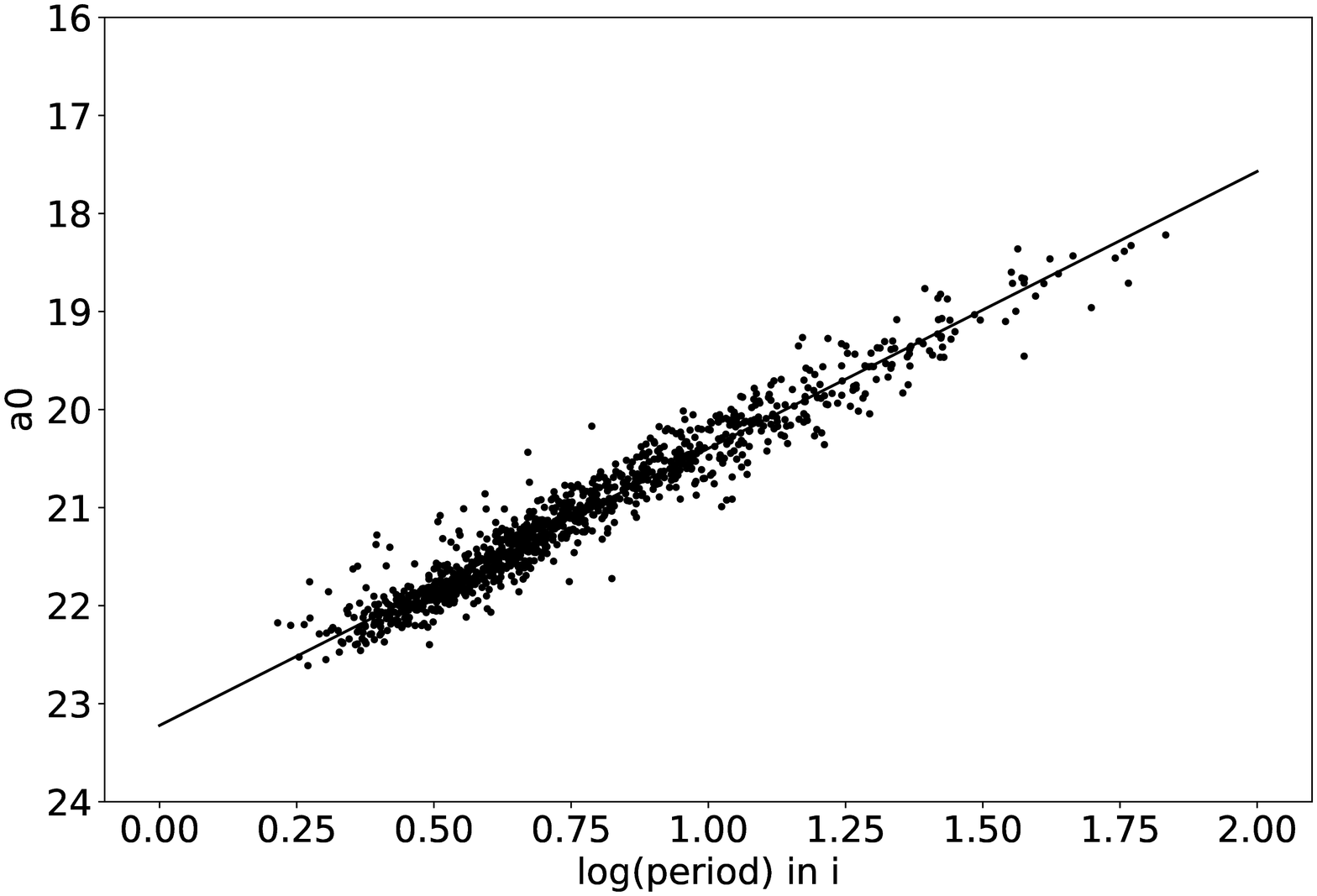}
    \includegraphics[width=0.5\linewidth]{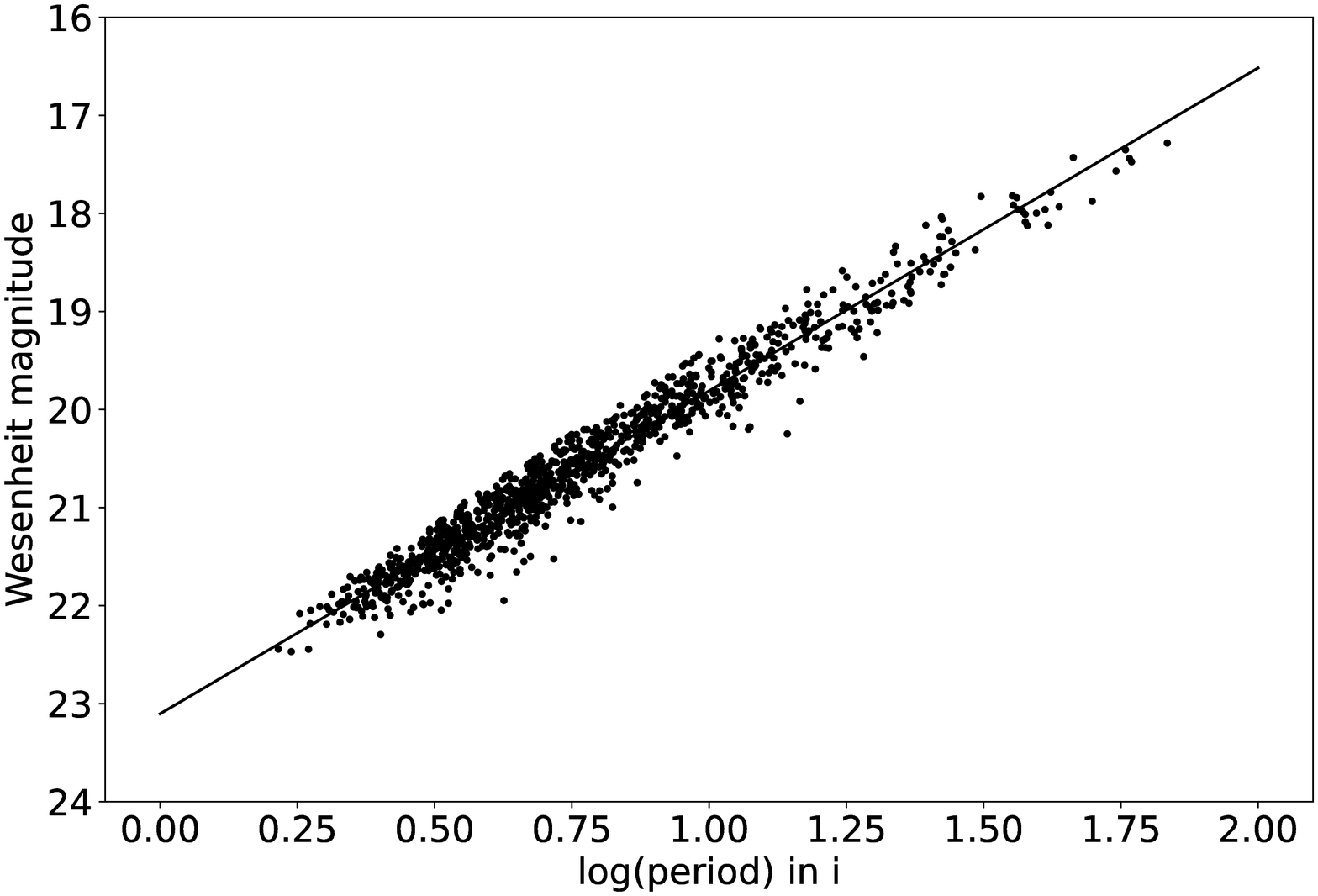}
\caption{We show the period-luminosity relationship and the best-fitted linear relation in the g, r, and i filters and for the extinction free Wesenheit magnitudes.}
\label{fig:10}
\end{figure}

Using the FM type Cepheids found in the previous section, we performed a linear fitting and derived the period-luminosity relation in the g, r, i filters and extinction free Wesenheit magnitudes.(see Fig.\ref{fig:10}). The four slopes and magnitudes obtained from the linear relation can be seen in Table \ref{tab:3}.

\begin{figure}[h!]
    \includegraphics[width=0.5\linewidth]{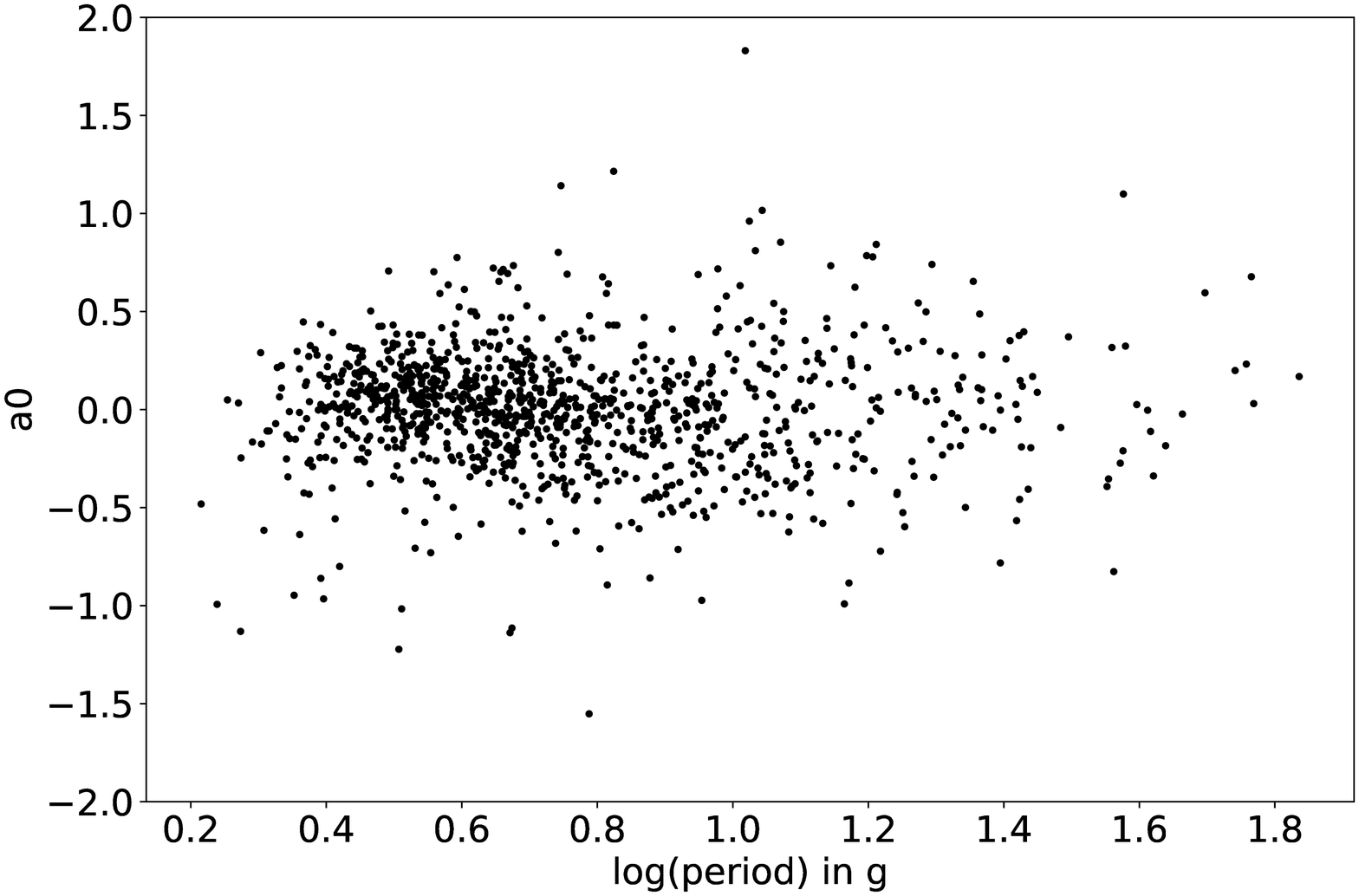}
    \includegraphics[width=0.5\linewidth]{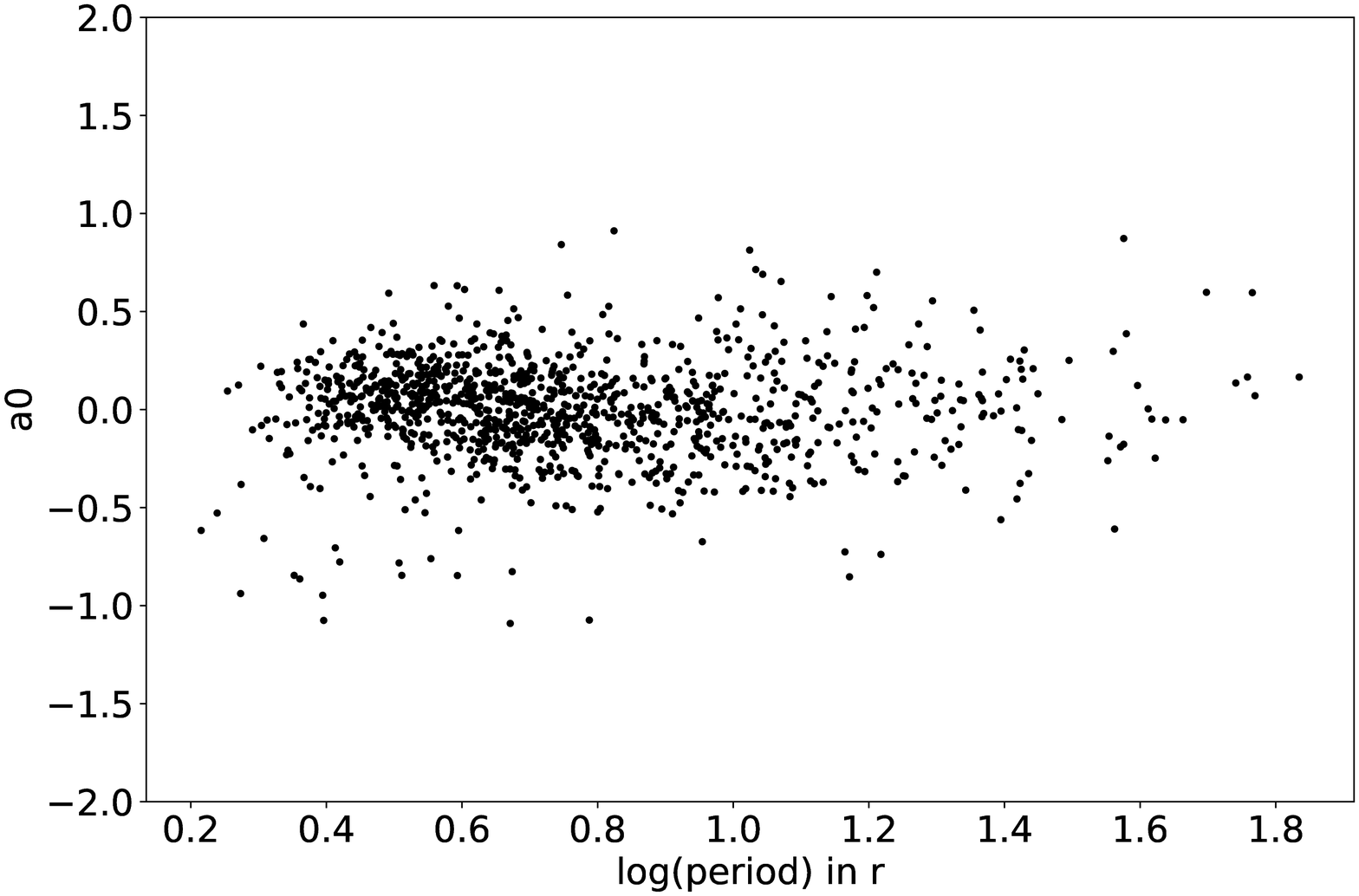}
    \includegraphics[width=0.5\linewidth]{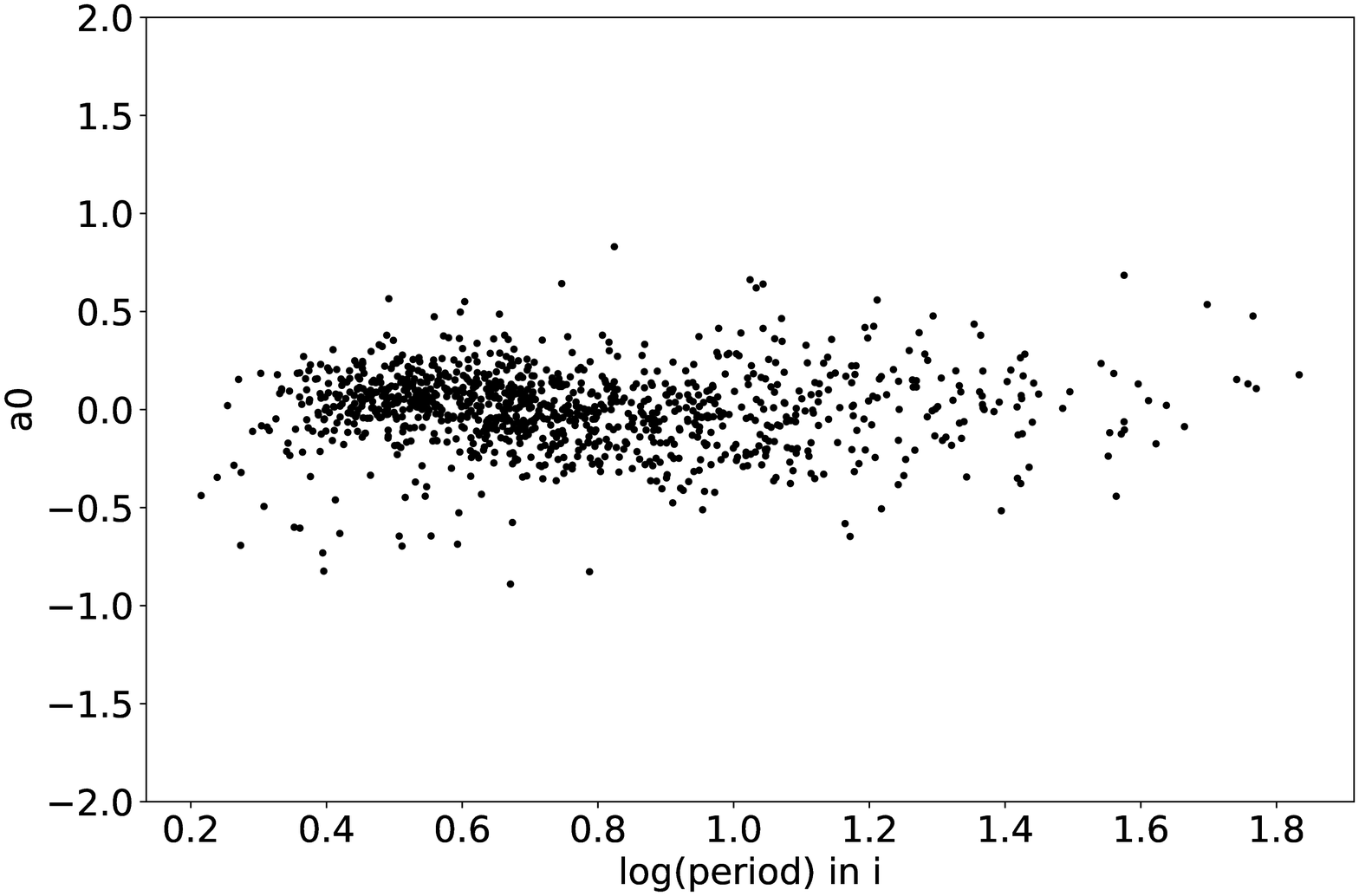}
    \includegraphics[width=0.5\linewidth]{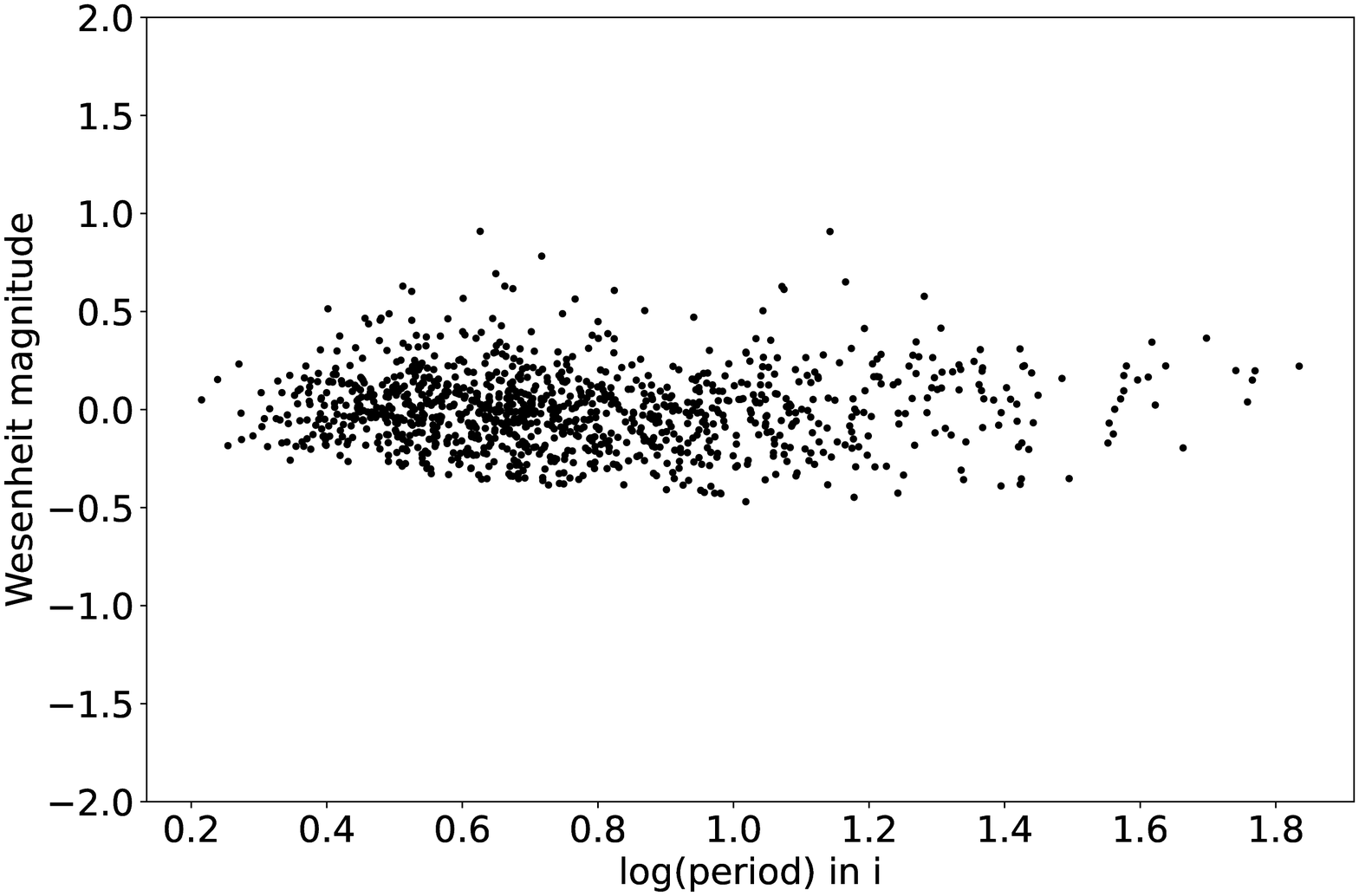}
\caption{We show the residual of period-luminosity relationship and the best-fitted linear relation in the g, r, and i filters and for the extinction free Wesenheit magnitudes.}
\label{fig:11}
\end{figure}

\setlength{\tabcolsep}{10pt}
\renewcommand{\arraystretch}{1}
\begin{table}[h!]
    \centering
    \begin{tabular}{|c|c|c|c|c|}
    \hline 
    filter & slope & magnitude & scatter($\sigma$) & period = 10 days \\ [2ex]
    \hline
    g filter & -2.319\textpm0.033 & 23.517\textpm0.027 & 0.310 & 21.198\textpm0.043 \\ [2ex]
    \hline
    r filter & -2.665\textpm0.027 & 23.266\textpm0.022 & 0.249 & 20.601\textpm0.035 \\ [2ex]
    \hline
    i filter & -2.826\textpm0.021 & 23.222\textpm0.017 & 0.199 & 20.396\textpm0.027 \\ [2ex]
    \hline
    Wesenheit Magnitude & -3.294\textpm0.021 & 23.103\textpm0.017 & 0.196 & 19.809\textpm0.027 \\ [2ex]
    \hline
    \end{tabular}
    \caption{Slopes and magnitudes of period-luminosity relation from FM Cepheids in g, r, and i filters and Wesenheit magnitudes}
    \label{tab:3}
\end{table}

We compared these zero points at period 10 days, where errors in the slope and errors in the zero point are largely decoupled, to the semi-empirical Cepheid period-luminosity relations in Sloan filter systems presented by \cite{2007ApJ...667...35N} for the LMC. \cite{2007ApJ...667...35N} derived a zero points of 14.647$\pm$0.045 in $g$, 14.208$\pm$0.034 in $r$, and in 14.104$\pm$0.029 in $i$. This translates into a difference of 6.551$\pm$0.062, 6.393$\pm$0.049, 6.292$\pm$0.040 mag between our M33 and the LMC zero points in the $g$, $r$, and $i$ filters, respectively. Adopting the 2\%-level distance estimate of LMC from late-type eclipsing binaries from \cite{2013Natur.495...76P}, we derived distance moduli $\mu$ of 25.044$\pm$0.083, 24.886$\pm$0.074, and 24.785$\pm$0.068 mag in the $g$, $r$, and $i$, respectively from our sample of M33 Cepheids.

With the M33 distance modulus in hand, we compare our results with previous results found by other studies. The first result we compare with is from an eclipsing binary by \cite{2006ApJ...652..313B}. \cite{2006ApJ...652..313B} presented a result of \textmu = 24.92 \textpm0.12 mag, using a detached eclipsing binary in M33 to determine the distance modulus. We also compare our results with other methods. For example \cite{2022ApJ...932...29L} used the spectra of blue supergiant stars to determine the distance modulus of M33. They obtained a distance modulus of \textmu = 24.93 \textpm 0.07 mag. Both of the eclipsing binary and blue supergiant methods are from early type stars and their results of distance estimate are in good agreement. \cite{2002AJ....123..244K} determined a distance modulus of \textmu = 24.81 \textpm0.04 based on the tip of the Red Giant Branch (RGB). The RGB stars are considered from older stellar population and the RGB method also agrees well with results from the early type stars. \cite{2004ApJ...614..167C} used the planetary nebula luminosity function and estimated a distance modulus of \textmu = $24.86^{+0.07}_{-0.11}$. \cite{2010AJ....140.1038P} derived a distance modulus of \textmu = 24.72 and 24.92 from red clump start in the $V$ and $I$-bands, respectively. Using Mira variables, \cite{2017AJ....153..170Y} derived a distance modulus of \textmu = 24.81$\pm$0.11. \cite{2010ApJ...724..799Y} obtained a distance modulus of 24.52 \textpm0.11 using RR Lyrae variables in M33. \cite{1987AJ.....93..833K} obtained a distance modulus of 24.64 \textpm 0.10 using various types of variable stars. 
Finally, we compare our results with previous Cepheid studies. \cite{2011ApJS..193...26P} determined a distance modulus of \textmu = 24.76 \textpm0.02 mag from a sample of 564 Cepheids. We are in good agreement with the results from \cite{2011ApJS..193...26P}. However, our results differ from \cite{2013ApJ...773...69G}, who obtained a distance modulus of 24.62$\pm$0.07 mag from 26 Cepheids. Our results also differ from \cite{2009MNRAS.396.1287S}, who derived a distance modulus of 24.53$\pm$0.11 from 167 fundamental mode Cepheids with the WIYN telescope. 
One possible explanation is that both our study and \cite{2011ApJS..193...26P} have engaged a larger sample compared to \cite{2009MNRAS.396.1287S} and \cite{2013ApJ...773...69G}. This suggest that the result from \cite{2009MNRAS.396.1287S} and \cite{2013ApJ...773...69G} might be subject to small number statistics. With our larger Cepheid sample, the Cepheid distance agrees well with other methods with larger distance estimate. 

\begin{figure}[h!]
\centering
\includegraphics[width = 14cm]{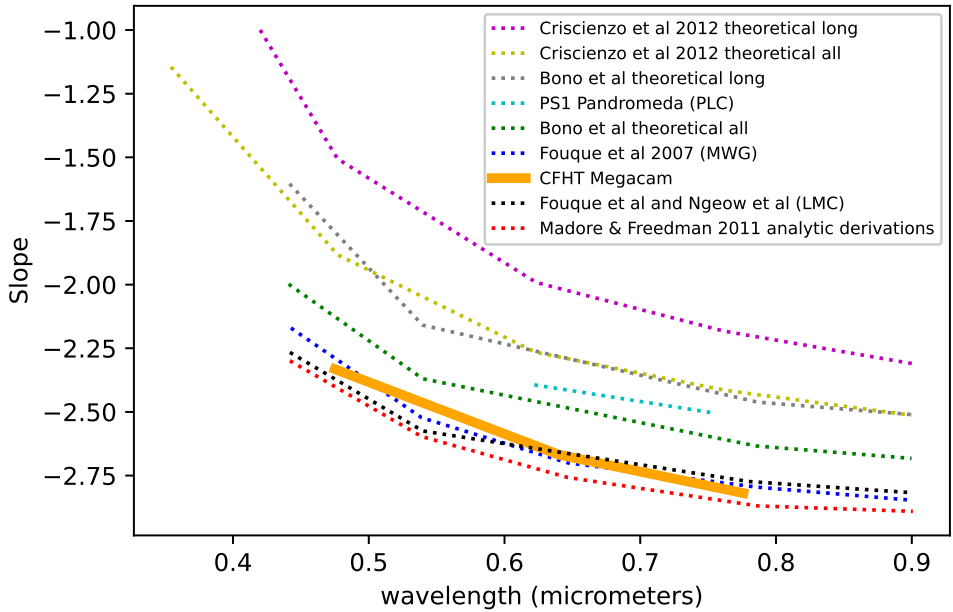}
\caption{The period-luminosity relation slopes we obtained plotted against the wavelength of the filter that corresponds to the slope (seen in magenta). The other lines are slopes from previous results.}
\label{fig:12}
\end{figure}

In addition to the distance estimate, we also compare the slope of period-luminosity relation v.s. wavelength of observations, as shown in figure \ref{fig:12}. The slopes we acquired agree well with theoretical predictions and the results from other Cepheids in the Milky Way and the Large Magellanic Cloud.

\section{Conclusion}
\label{sec.conclusion}
We used archived data of M33 variable sources from a CFHT/Megacam survey carried out by \cite{2006MNRAS.371.1405H} to study the properties of Cepheids in the M33 galaxy. We started with more than 36,000 variable sources and identified a sample of 1,622 Cepheids from period search, visual classification, and locating the instability strip domain from Wesenheit color cuts. We classified the Cepheids into FM type, FO type, or type II Cepheids using the amplitude ratio and visual classification. 
With a clean sample of FM type Cepheids, we derived the period-luminosity relations in g, r, i filters, and Wesenheit magnitudes. 
By comparing with the results from semi-empirical Cepheid period-luminosity relation from LMC in similar filters, we obtain distances to M33 comparable with other distance indicators such as eclipsing binaries, blue super giants, and red giant branch stars. This suggests that discrepancies in distance estimates from previous Cepheid studies might originate from small number statistics.

\begin{acknowledgments}
We thank the reviewer for the insightful comments that improved the manuscript. Samuel Adair is supported by the Akami internship program.
This work is supported by JSPS KAKENHI Grant 17K14256.

This work is based on observations obtained with MegaPrime/MegaCam, a joint project of CFHT and CEA/DAPNIA, at the Canada-France-Hawaii Telescope (CFHT) which is operated by the National Research Council (NRC) of Canada, the Institut National des Science de l'Univers of the Centre National de la Recherche Scientifique (CNRS) of France, and the University of Hawaii. The observations at the Canada-France-Hawaii Telescope were performed with care and respect from the summit of Maunakea which is a significant cultural and historic site.

\end{acknowledgments}

{}



\end{document}